\documentclass[preprint,12pt,3p]{elsarticle}

\let\today\relax
\makeatletter
\def\ps@pprintTitle{%
    \let\@oddhead\@empty
    \let\@evenhead\@empty
    \def\@oddfoot{\footnotesize\itshape
         {Preprint submitted to \texttt{Sensors}} \hfill\today}%
    \let\@evenfoot\@oddfoot
    }
\makeatother

\usepackage{tabularx}
\usepackage{subfig}
\usepackage{placeins}

\begin{document}
%%%%%%%%%%%%%%%%%%%%%%%%%%%%%%%%%%%%%%%%%%

\begin{frontmatter}

\title{Staying Alive -- CPR Quality Parameters from Wrist-worn Inertial Sensor Data with Evolutionary Fitted Sinusoidal Models}

\author[label2]{Christian Lins\corref{cor1}} %\fnref{label3}
\address[label1]{Carl von Ossietzky University Oldenburg, Dep. Health Services Research, \\Ammerl\"ander Heerstr. 140, 26129 Oldenburg, Germany}
\address[label2]{OFFIS - Institute for Information Technology, Div. Health, \\Escherweg 2, 26121 Oldenburg, Germany\fnref{label4}}

\cortext[cor1]{Corresponding author}

\ead{christian.lins@offis.de}
%\ead[url]{author-one-homepage.com}

\author[label1]{Andreas Klausen}
%\ead{andreas.klausen@uni-oldenburg.de}

\author[label1]{Sandra Hellmers}
%\ead{sandra.hellmers@uni-oldenburg.de}

\author[label1,label2]{Andreas Hein}
%\ead{andreas.hein@uni-oldenburg.de}

\author[label1]{Sebastian Fudickar}
%\ead{sebastian.fudickar@uni-oldenburg.de}

\begin{abstract}
In this paper, a robust sinusoidal model fitting method based on the Differential Evolution (DE) algorithm for determining cardiopulmonary resuscitation (CPR) quality-parameters – naming chest compression frequency and depth – as measured by an inertial sensor placed at the wrist is presented. Once included into a smartphone or smartwatch app, the proposed algorithm will enable bystanders to improve CPR (as part of a continuous closed-loop support-system). By evaluating the precision of the model with data recorded by a Laerdal Resusci Anne mannequin as reference standard, a low variance for compression frequency of $\pm 2.0$ cpm has been found for the sensor placed at the wrist, making this previously unconsidered position a suitable alternative to the typical placement in the hand for CPR-training smartphone apps.
\end{abstract}


\end{frontmatter}

\section{Introduction}
\noindent
Sudden cardiac arrests (SCA) is one of the most prominent diseases (350,000-700,000 individuals a year in Europe are affected \cite{Berdowski2010, Grasner2011, Grasner2013}). SCA can significantly affect the independent living of the victims if medical treatment is not available within a few minutes \cite{Perkins2015,DeMaio2003}. In case of a cardiac arrest, the transport of oxygen and glucose to the cells of the human body stops immediately due to the disrupted heart function. This results in irreparably cell damage if the blood circulation is not re-established, e.g. via cardiopulmonary resuscitation (CPR). For the cells of the nervous system including the brain, that means that the functionality reduces after 10 seconds (i.e. loss of consciousness) \cite{Schmidt2011}. The death of the cells begins after about 3 minutes \cite{Schmidt2011}.

Medical personnel such as paramedics are trained in Advanced Life Support (ALS) \cite{Soar2015} methodology that includes CPR. Unfortunately, paramedics are usually not immediately available if a cardiac arrest occurs in the field. With the typical median response-time of paramedics being about 5-8 min \cite{Perkins2015} and a decreasing likelihood of survival with every minute without CPR, victims depend on initial CPR-support of non-specialist bystanders within the first “golden” minutes after a cardiac arrest to prevent negative long-term effects. Since these bystanders can offer essential initial resuscitation support, corresponding technical solutions to support them with online feedback regarding the quality of CPR are required. 

For the following discussion, the optimal Basic Life Support (BLS) \cite{Perkins2015} procedure (the methodology aiming for non-specialists) is worthwhile to be briefly recapitulated: In case of a cardiac arrest, it is essential to ensure sufficient oxygenation of the nerve cells via correctly conducted cardiac massages (chest compressions) as the most critical countermeasure. During this cardiac massage, the heart is compressed by orthogonal pressure onto the breastbone. In order to sustain a minimal blood circulation to carry oxygen to the nerve cells, a chest compression frequency (CCF) of the cardiac massage should range from 110 $\pm10$ compressions per minute (cpm), and a chest compression depth (CCD) of approximately $5.5 \pm0.5$ cm is required. Ideally (but not necessarily) the procedure is combined with rescue breathing, to improve the chance of survival and reduce neurological deficits \cite{Perkins2015}.

While typical bystanders can develop a sufficient feeling of semi-ideal compression depth, the constant application of the correct compression-frequency and -depth is challenging, especially for extended periods of cardiac massage with the associated muscle-fatigue and mental pressure. Thus, instant feedback regarding the correct execution (regarding CCD and CCF) during cardiac arrest will be beneficial for untrained bystanders. Such feedback could be derived from monitoring the quality of the cardiac massage online from the vertical acceleration as measured by inertial measurement units (IMUs) and giving continuous feedback (and adjustment hints) regarding CCD and CCF. Corresponding smartphone applications have been shown to be well suited for this purpose due to their high availability \cite{Kalz2014, Ahn2016}. Their general benefit regarding CCF/CCD has been confirmed by Renshaw et al. for the BHF PocketCPR who recognised an improved performance (from 66 to 91 cpm) and increased confidence of bystanders \cite{Renshaw2017}. However, for such CPR training apps, a highly accurate CPR information (regarding CCF and CCD) is an essential requirement \cite{Ahn2016} which is achieved partially by the existing implementations (as summarized in Table \ref{tab:cpr_training_ref}).

\begin{table}[h]
    \caption{Accuracies of existing chest-compression algorithms.}
    \label{tab:cpr_training_ref}
    \centering
    \small
    \begin{tabularx}{\linewidth}{p{1.9cm}|p{4.9cm}|p{1.5cm}|p{1.5cm}|p{2.1cm}|p{1.6cm}}
        Measurement Position & Algorithm & CCF Error [cpm] & CCD Error
[mm] & Reference System & Year\\\hline

On chest &	Spectral techniques on short acceleration intervals &	< 1.5 &	< 2 &	photoelectric distance sensor &	2014 \cite{Gonzalez-Otero2014}\\

On chest &	Butterworth HP filter, 2x integration, manual reset	& - &	1.6 (within 95\%) &
	mannequin potentiometer &
	2002 \cite{Aase2002}\\

On chest &	Weighted smoothing, double 2x (transient component emphasizing + integration), peak detection (U-CPR)&
	- &	1.43 (1.04) mean &	mannequin potentiometer &
	2015 \cite{Song2015}\\

On chest&	PocketCPR&	- &	1.01 (0.74) mean&	mannequin potentiometer &
	2015 \cite{Song2015}\\

On chest&	Spectral analysis of acceleration&	0.9 median&	1.3 median &	displacement sensor & 2016 \cite{DeGauna2016}
    \end{tabularx}
\end{table}

Most of these approaches use Fast Fourier Transformation (FFT) to determine the regular frequencies from the inertial data and identifying the main frequencies from the frequency spectrum via peak detection: While being straightforward, this approach is susceptible for erroneous peak selection and a resulting frequency shift of the CCFs by orders of magnitudes. 
Also, the double-integration of the acceleration signal – a common processing step for the determination of the displacement-vector as a preprocessing step for the CCD is challenging due to the signal drift if the accelerometer is not perfectly aligned with the gravity axis \cite{ladetto2000foot}.
Consequently, the typical integration process is inherently unstable and leads to relevant errors unless boundary conditions are applied for each compression cycle (e.g. in very short windows) \cite{DeGauna2016}. 

In contrast, the use of a robust sinusoid model, i.e. sine curve, might cover such phenomena and its subsequent effects on the derived CPR parameters more robustly, due to its implicit periodic accordance with the CPR. A concept that has successfully proven itself for depth image based motion capture of CPR movements \cite{Lins2018, Lins2018arx}, but still needs to be confirmed for use with acceleration data recorded directly via IMUs at rescuers.  

Furthermore, the discussed algorithms have been mainly evaluated for the use of IMUs in a grasp-in-hand use, a position that has been reported to be a rather uncomfortable and disturbing positioning for lay-persons \cite{Park2016}, which as well might mislead them into learning incorrect postures \cite{Park2017}. To overcome this drawback, Park et al. \cite{Park2016} proposed the fixation of smartphones via an armband on the dorsum manus or at the arm and had shown an increased convenience in comparison to the common grasp-in-hand approach. However, they reported a reduced sensitivity, which they explained with the amplified inertial forces resulting from the additional device’s swing. Similarly Ruiz de Gauna et al. \cite{Gauna2015} compared sensor placement at the dorsum manus with one fixed to the forearm 7 cm above the wrist and confirmed a significantly increased error for the forearm placement with median errors of 3.1 mm (1.4–5.1) and 9.5 mm (6.8–12.9) for the dorsum manus and for the forearm, respectively. 

Consequently, while IMUs have generally proven to be a precise and practical approach to measuring CC depth and frequency during CPR, user-friendliness via smartphones is a challenging task as it affects the quality of CPR for bystanders. In contrast, smart\emph{watches} hold benefits over the use of smartphones regarding usability and could be expected to achieve higher reliability towards arm-movements, since being potentially less affected by hand movements. Furthermore, they overcome challenging aspects of reduced tactile pressure sensation at the hands. However, the use of IMUs on alternative placements was repetitively found challenging for sufficient accuracy, and the suitability of smartwatches 
for CPR training and online-support regarding the sensitivity of CCD and CCF detection yet has to be investigated. 
Consequently, with the article at hand, we aim to investigate the following two research questions: 

\begin{enumerate}
    \item How suitable are wrist-worn inertial sensors (e.g. smartwatches) for the online detection of the chest compression during CPR regarding the accuracy of resulting CCF and CCD parameters in comparison to both the Resusci Anne training mannequin and the typical hand-holding as gold standard?
    
    \item How suitable is the DE algorithm for fitting a sinusoidal model of the chest compression during CPR regarding the accuracy of resulting CCF and CCD parameters for the considered sensor placements in comparison to both the Resusci Anne training mannequin as reference system?
\end{enumerate}

Thereby, in Section 2 our system approach for inertial data, the study design and applied evaluation methodology are introduced. In section 3 the results of the study are discussed. The article is concluded 
in Section 4. 

\section{Materials and Methods}

\subsection{System}

\begin{figure}[h]
    \centering
    \includegraphics[width=0.9\textwidth]{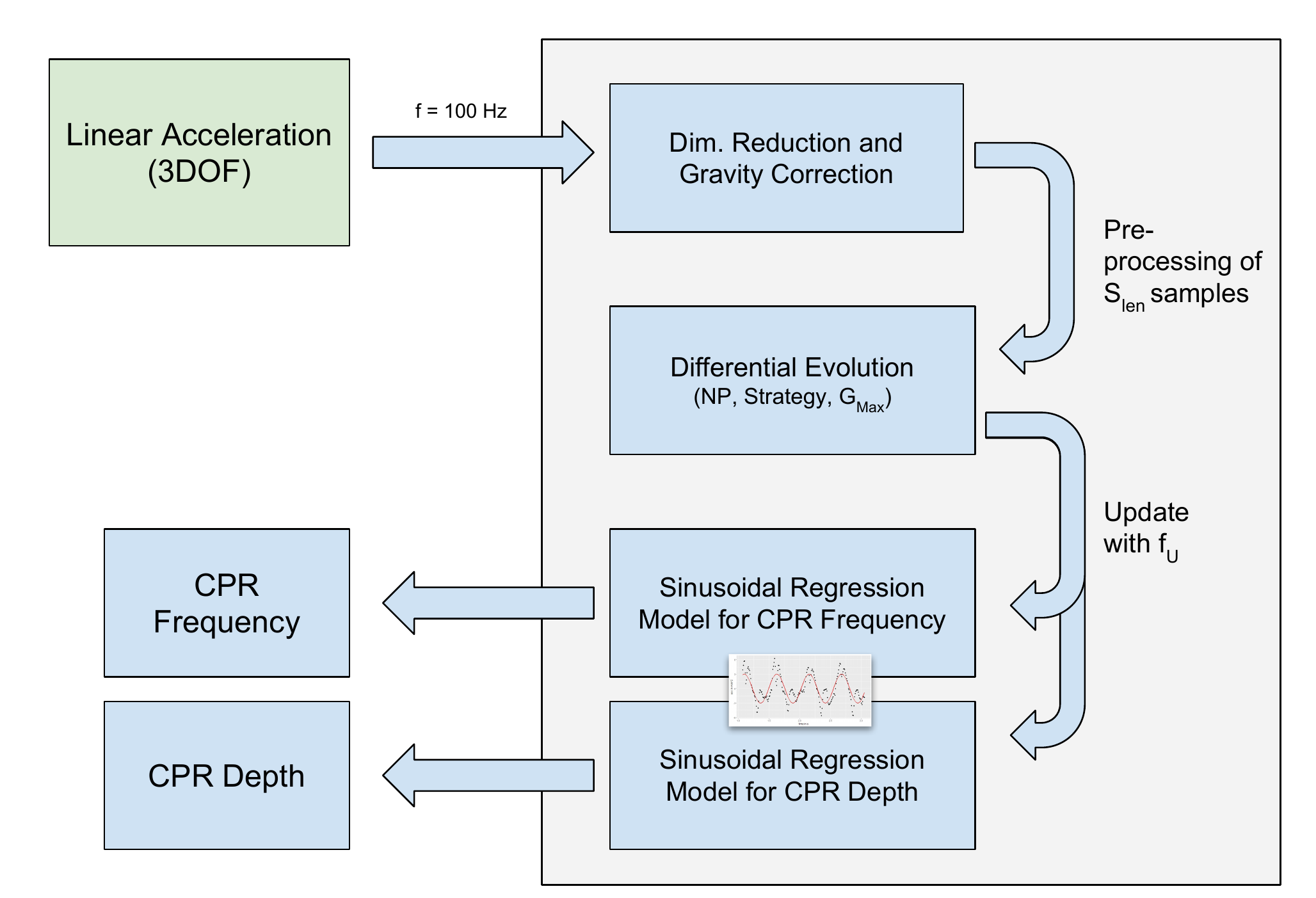}
    \caption{System concept overview of the proposed method}
    \label{fig:system_concept}
\end{figure}
\noindent
The system approach presented here (Figure \ref{fig:system_concept}) uses data from a wrist-worn inertial 
accelerometer. The sensor data is converted from the usual representation in 3DOF with gravity to the absolute acceleration without gravity, i.e. the length of the acceleration vector, with only one dimension.

The acceleration is measured in three dimensions and always includes gravity. Thus, the gravity must be subtracted from the accelerometer signal:

\begin{equation}
    a=\sqrt{\left({a_x}^2+{a_y}^2+{a_z}^2\right)} - 9.81 \frac{m}{s^2}
\end{equation}

These one-dimensional values over time are window-based (window length $S_{len}$) adapted to a sinusoidal functional model. The evolutionary Differential Evolution (DE) algorithm is used for this purpose. The fitting produces adapted models with every ${f_U}^{-1}$ seconds, from which the CPR parameters frequency (CCF) and compression depth (CCD) can be derived.

\subsection{DE fitting of accelerometer data to model}
\begin{figure}[h]
    \centering
    \includegraphics[width=\textwidth]{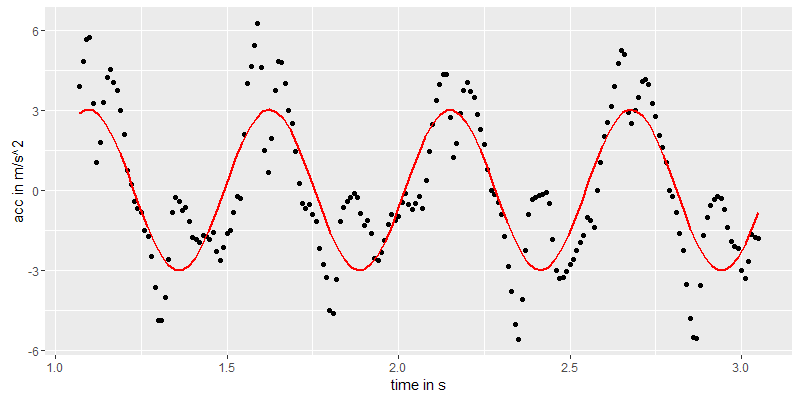}
    \caption{Fitting of the sinusoid model to the accelerometer data.}
    \label{fig:fitting}
\end{figure}

\noindent
The proposed approach utilizes the periodic nature of the CPR to fit the signal of the accelerometer to a sine curve (see Figure \ref{fig:fitting}). The generic parameterized sine function can be written as follows:
\begin{equation}
    \hat{y}\left(t\right)=A\cdot sin\left(\omega t+\rho\right)+D
    \label{eq:sine1}
\end{equation}

\noindent
Parameter A and $\omega$ are of primary interest here: $A$ is the amplitude, $\omega$ the angular frequency ($D$ is the horizontal displacement, $\rho$ the phase shift). Assuming that the arms of a person performing CPR are orthogonal (and rigid) on the patient's chest, the relative movements of the rescuer's arms are equal to the chest compression depth. Additionally, the frequency of low to high to low compression depth represents one compression cycle.
Unfortunately, it is not applicable to fit the accelerometer data directly to the sine curve (see Equation 1) and derive the displacement of the arm from it. The accelerometer measures -- hence the name -- the acceleration, which must be integrated twice  to determine the vertical displacement ($\int\int$ acceleration $\rightarrow \int$ velocity $\rightarrow$ displacement). The double integration of the acceleration values induces errors if the sensors are not perfectly calibrated, so this is rarely a practicable way. To avoid this problem, we use the second derivative of the sine function as a model function. Thus, the analytical solution of the double integration is already known ($\hat{y}$, Equation \ref{eq:sine1}):

\begin{equation}
    y\left(t\right)=\frac{\Delta}{\Delta t^2}A\cdot sin\left(\omega t+\rho\right)+D
\end{equation}

\begin{equation}
    y\left(t\right)=-A\omega^2\cdot sin\left(\omega t+\rho\right)
    \label{eq:sin_function}
\end{equation}

On a fitted function, parameter $\omega$ is the CPR frequency, $2\cdot A$ is the compression depth. To fit the function $y\left(t\right)$, we minimize the root mean squared errors (RMSE) using an evolutionary approach. Thus, we formulate the minimization problem as follows:

\begin{equation}
    min\sqrt{\frac{1}{\left|a\right|}\sum_{t=0}^{\left|a\right|}\left(a\left(t\right)-y\left(t\right)\right)^2}
    \label{eq:minimization_problem}
\end{equation}

In the equation, $a$ is the vector of accelerometer measurements. 

A common evolutionary algorithm is the Differential Evolution \cite{Storn1997} that works particularly well with nonlinear, i.e. sinusoidal cost functions. DE searches and evaluates a parameter space in parallel and finds multiple near-optimal but distinct solutions to a problem. The DE algorithm is used here to continously solve the three variables $\omega, A, \rho$ of a sinusoidal curve (Equation \ref{eq:sin_function}). As all evolutionary algorithms, DE is population-based and optimizes the population throughout several generations:

\begin{equation}
    x_{i,G}\mathrm{\ with\ }i=1..NP,G=1..G_{max}
\end{equation}

$x_{i,G}$ is a 3-dimensional vector of individual $i$ for generation $G$. So in every generation NP individuals are optimized up to $G_{max}$ generations. One individual represents one possible solution to the minimization problem (see Equation \ref{eq:minimization_problem}). The optimization is done between a transition from one generation $G$ to another generation $G+1$. Within each transition from one generation to the following DE comprise the steps mutation, crossover and selection.

\subsubsection{Mutation}
\noindent
For every generation, a mutation step is performed for every individual $x_{i,G}$. We used the default \texttt{DE/rand/1/bin} step with a fixed amplification factor $F=0.8$ \cite{Storn1997,Mullen2011,Peterson2016}: 

\begin{equation}
    v_{i,G+1}=x_{r_1,G}+F\cdot\left(x_{r_{2,G}}-x_{r_{3,G}}\right)
    \label{eq:mutated_individual}
\end{equation}

In Equation \ref{eq:mutated_individual} $v$ is the mutated individual and $r_1,r_2,r_3\in\{1,2,\ldots,NP\},r_1\neq r_2\neq r_3\neq i$ randomly chosen.

\subsubsection{Crossover}
\noindent
The crossover step determines which of the four parameters per individual are preserved in the next generation. For every parameter, a uniform random number $r\in\left[0,1\right]$ is chosen. If $r\ \le CR\ =\ 0.5$ then the parameter from the mutant is chosen, otherwise the one from the original individual is continued with.

\subsubsection{Selection}
\noindent
The selection step decides which individuals are passed to the next generation by evaluating them against the cost function. Herein, the RMSE are summed up for every solution candidate $x_{i,G}$ as cost function:

\begin{equation}
    \sum_{t=0}^{T}\left(a\left(t\right)-y_{x_{i,G}}\left(t\right)\right)^2
\end{equation}

with $a$ being a $T$-length vector of samples (accelerometer data) and $y_{x_{i,G}}$ the parameterized sinusoid function of individual $x_{i,G}$.
Once the DE optimization is finalized, i.e. when $G_{max}$ was reached, the individual with the lowest RMSE represent the parameter of the sinusoid model. From these parameters, the CCF and CCD can be obtained with $CCF=\frac{\left|\omega\right|}{\pi} \cdot 60$ cpm and $CCD=\left|2A\right|$ m.

\subsection{Experimental setup}
\noindent
The Laerdal Resusci Anne Simulator mannequin was used as a reference system and was placed on the floor (see Figure 3). Within the Resusci Anne simulator, sensors measure the depth of thorax compression and decompression, the frequency of the compressions and the volume of ventilation.

\begin{figure}[h]
    \centering
    \includegraphics[width=\textwidth]{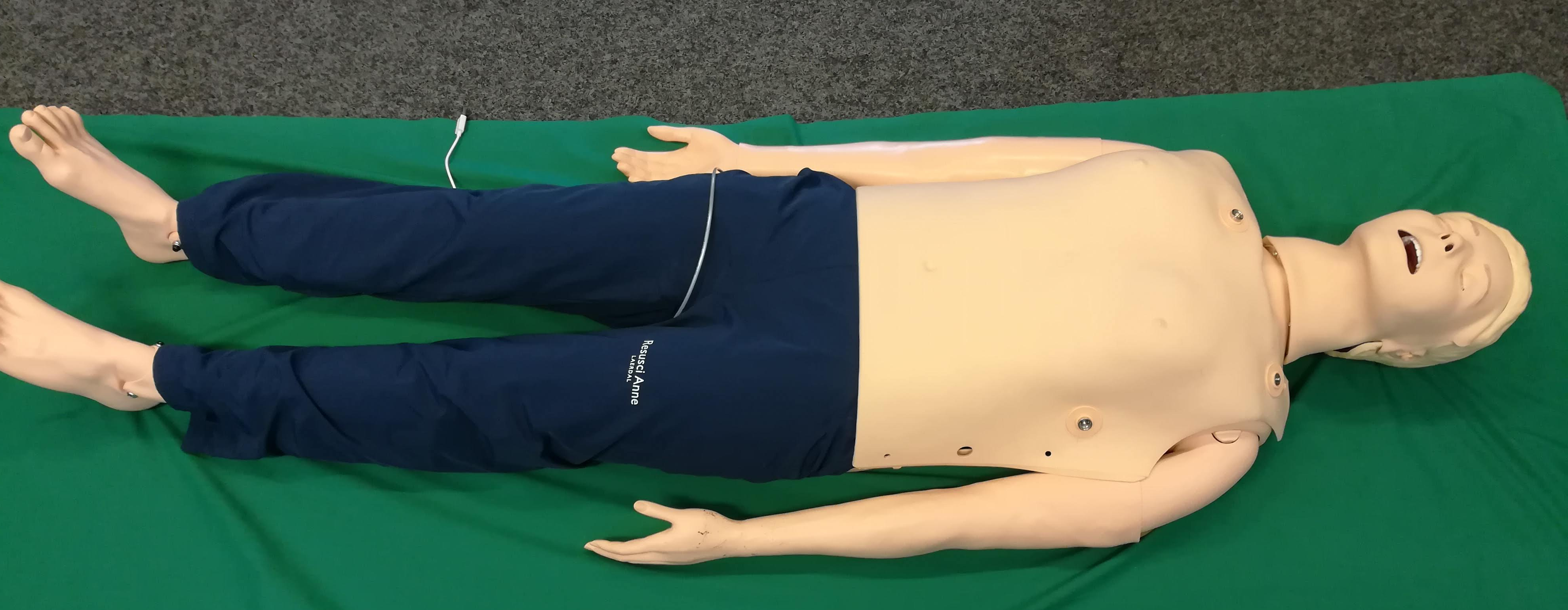}
    \caption{Resusci Anne training mannequin.}
    \label{fig:resusci_anne}
\end{figure}

\begin{figure}[h]
    \centering
    \includegraphics[height=3cm]{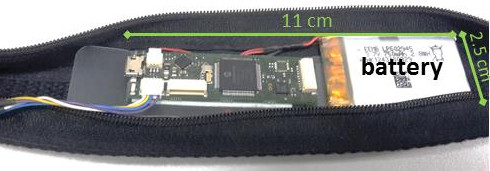}
    \includegraphics[height=3cm]{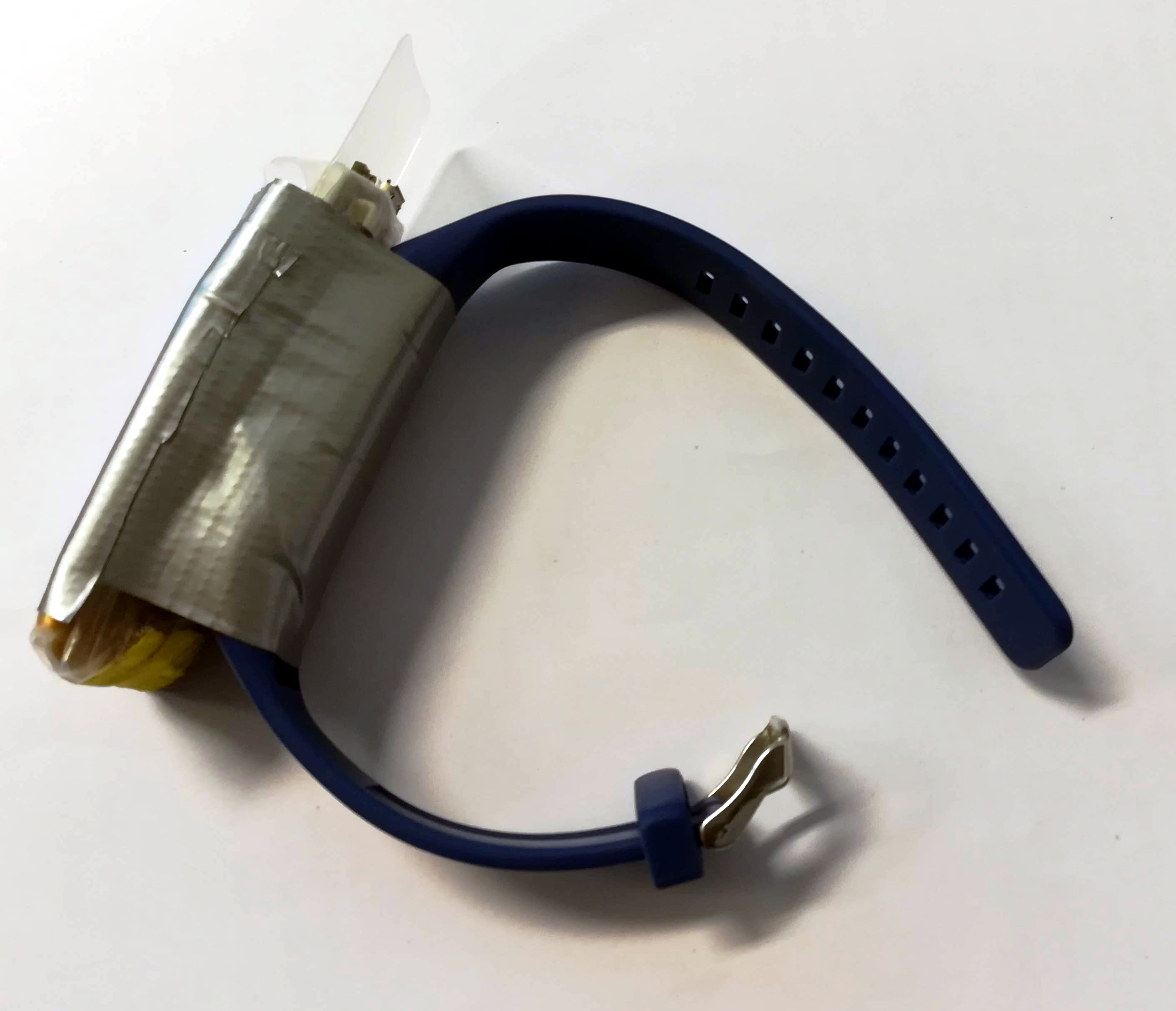}
    \caption{IMU-Sensor used in the study unpacked (left) and with bracelet (right).}
    \label{fig:sensor}
\end{figure}

\noindent
In addition, two IMU sensors have been applied (see Figure \ref{fig:sensor}), one was placed at the left wrist of the participant with a bracelet, and the other one was placed between the hands of the participant (between the back of the hand of the first hand and the palm of the second hand). The participants were asked to perform CPR compressions on the mannequin with standard CPR frequency and depth for about 120 seconds. The sensor, as well as the Resusci Anne, were collecting data which was synchronized manually after the recording using visual plots of the accelerometer signal.
 
\noindent
The 9DOF-IMU sensor contains a Bosch BMA180 triaxial accelerometer with sensitivity ranges from 1G up to 16G and sampling rates up to 1200 Hz. A reduced sampling rate of 100 Hz was used to reduce the amount of data for the system.

The participants were asked to place themselves on any side of the training mannequin and to perform the CPR on the training mannequin at their own discretion.  During the resuscitation session, the training mannequin recorded the compression movements with its internal sensors. The recording ended after two minutes.

\subsection{Evaluation processing steps}
\noindent
The DE implementation of the Python package scipy/1.2.0 \cite{scipy} was used on Python/3.6 for data processing and the evaluation. 

\begin{figure}[h]
    \centering
    \includegraphics{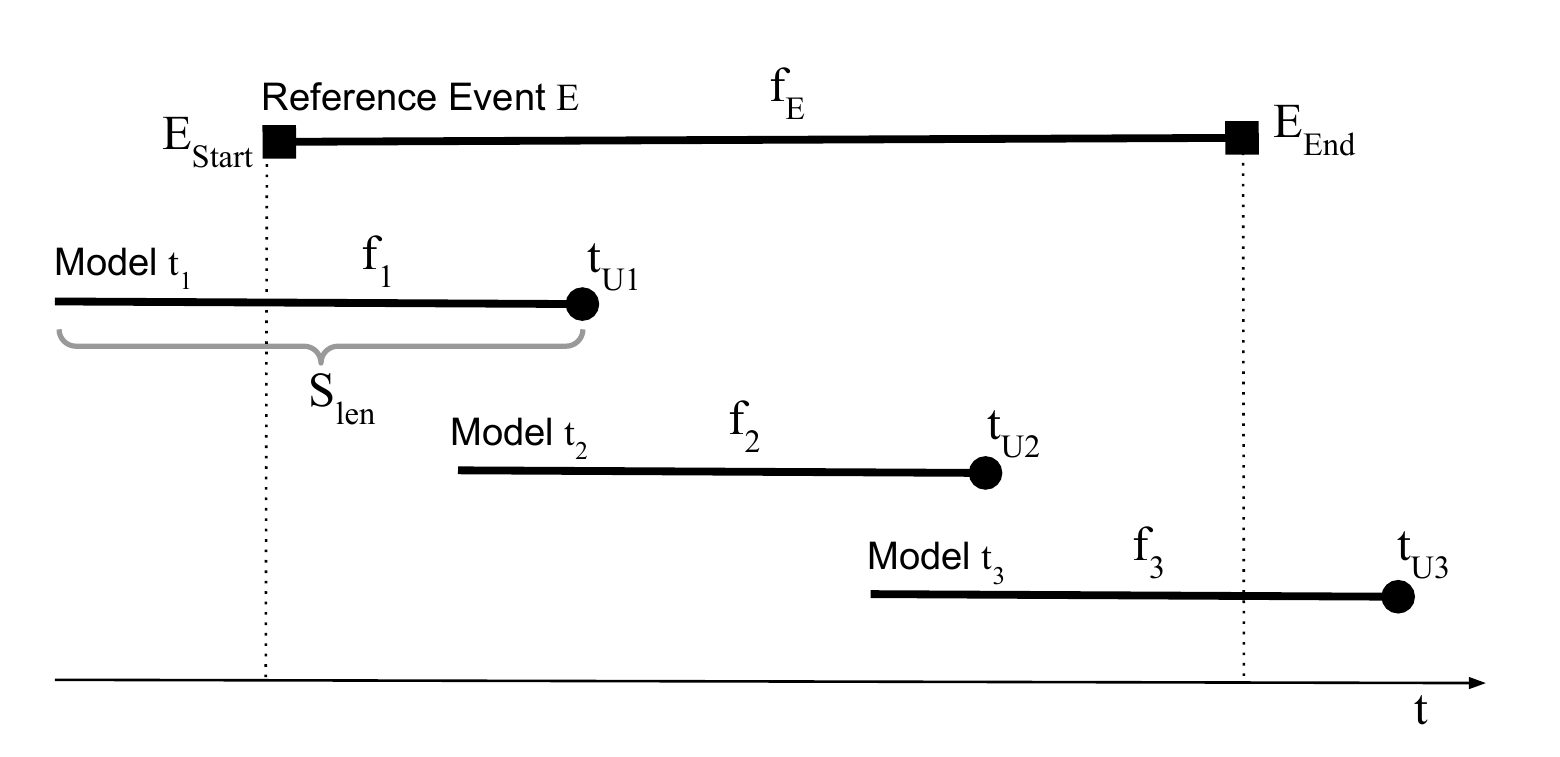}
    \caption{How model predictions and reference values are compared.}
    \label{fig:model_comparison}
\end{figure}

For every compression cycle recognized by the reference system, a 3-second window ($S_{Len} = 3s$) of the sensor data is used to fit the sinus model every second ($f_U = 1s^{-1}$) \cite{Lins2018arx}.
Here, it is $S_{len} \leq f^{-1}_{U}$, so the used datasets $S$ are interleaved (see Figure \ref{fig:model_comparison}). 

Also, one Resusci Anne event $E$ may be smaller, equal, or larger than $f^{-1}_{U}$ so that we must combine one or more model predictions before comparing it with $E$ (Equation \ref{eq:pred_combine}).
Thereby, the weighted mean of $n$ subsequent model predictions within each interval $(E_{Start}, E_{End})$ is calculated with the overlap ratio $\sigma$ representing the weight:

\begin{equation}
\label{eq:pred_combine}
    p(t) = \frac{1}{\sum_{i=1}^{n} \sigma_i} \sum_{i=1}^{n} \sigma_i f_i
\end{equation}

For each compression event $E$, a corresponding prediction $p$ is determined according to Equation \ref{eq:pred_combine}. The error is then the absolute difference between prediction CCF/CCD and reference CCF/CCD. All errors are then combined using the median.

\FloatBarrier
\newpage
%%%%%%%%%%%%%%%%%%%%%%%%%%%%%%%%%%%%%%%%%%
\section{Results and Discussion}
\noindent
We performed the study with 15 participants, aged 21-48 (median 33), 11 male, 4 female. The participants were recruited within students and staff of University of Oldenburg. 
%The study received an ethics approval no. “Drs. 24/2017” of the ethics committee of the University of Oldenburg.

\subsection{Model Prediction Results}
\noindent
We obtained sensor data from two different locations (between hands, and on the wrist), used this data to fit sinusoidal models and compared the predictions of this models to the internal sensor of the Resusci Anne reference system. Table \ref{tab:results} lists the main results comparing the frequency and depth prediction errors for the two sensor locations.

\begin{table}[h]
    \centering
    \caption{Median prediction error over all participants with system parameters $S_{Len} = 3s$ and $f_U = 1s^{-1}$.}
    \begin{tabular}{l|c|c}
Sensor Location & CCF Median Error & CCD Median Error \\\hline
Hand  &	2.1 cpm & 0.5 cm \\
Wrist &	2.0 cpm & 1.0 cm
    \end{tabular}
    \label{tab:results}
\end{table}

\begin{figure}[h]
    \centering
    \includegraphics[width=7.75cm]{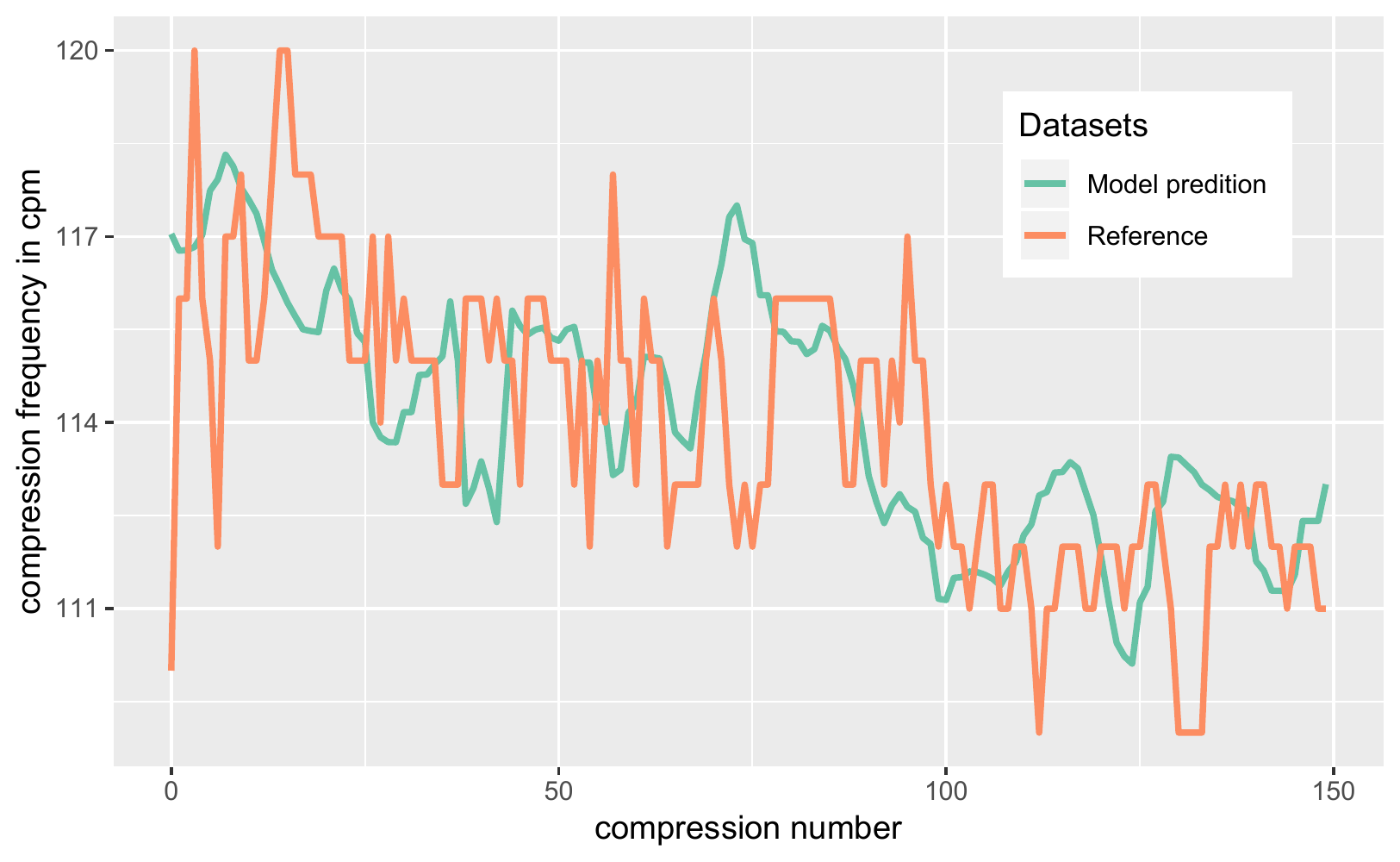}
    \includegraphics[width=7.75cm]{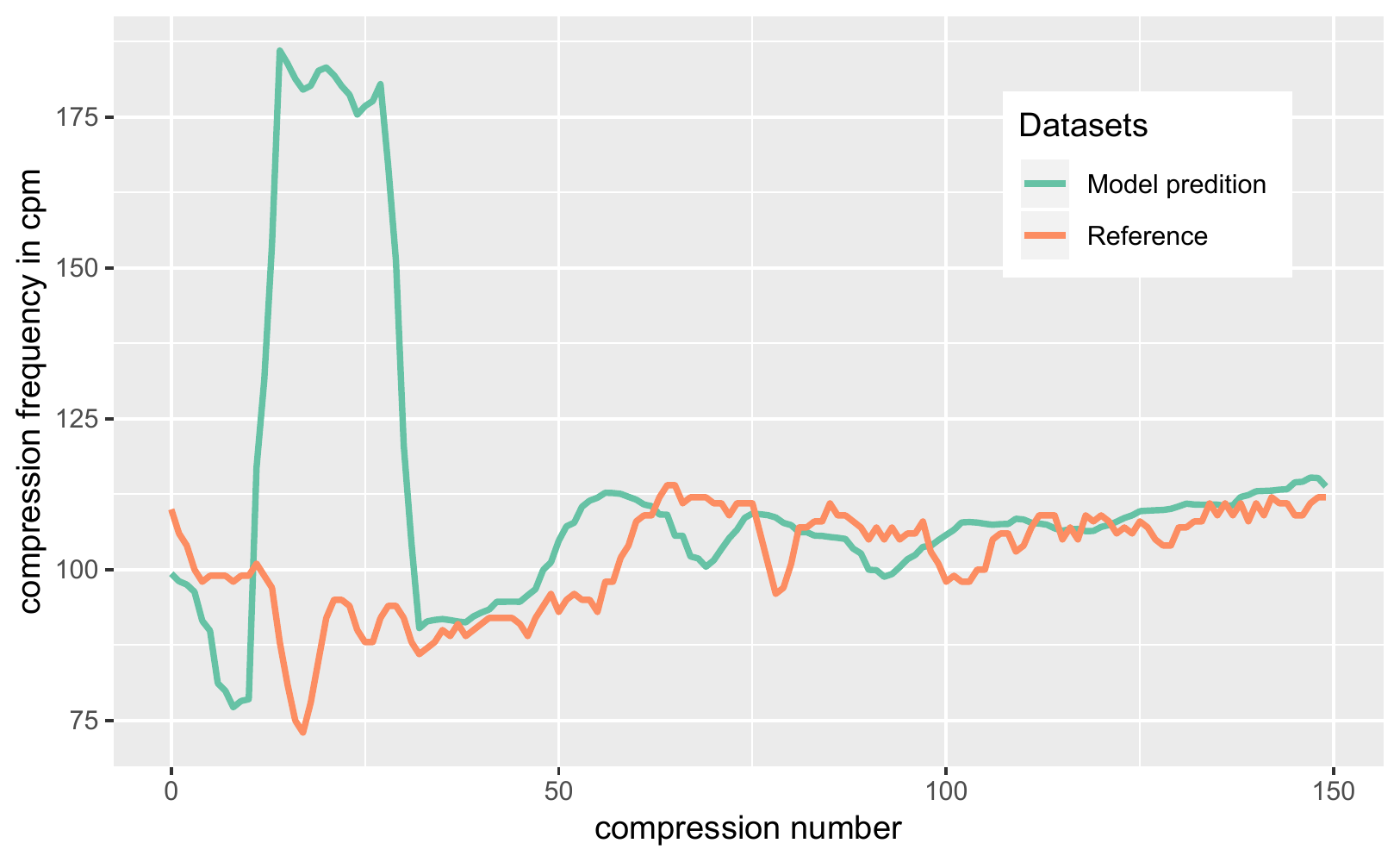}
    \caption{Wrist-worn sensor best-case (left, participant 1) and worst-case (right, participant 10) frequency prediction.}
    \label{fig:results}
\end{figure}

\begin{figure}[h]
    \centering
    \includegraphics[width=7.75cm]{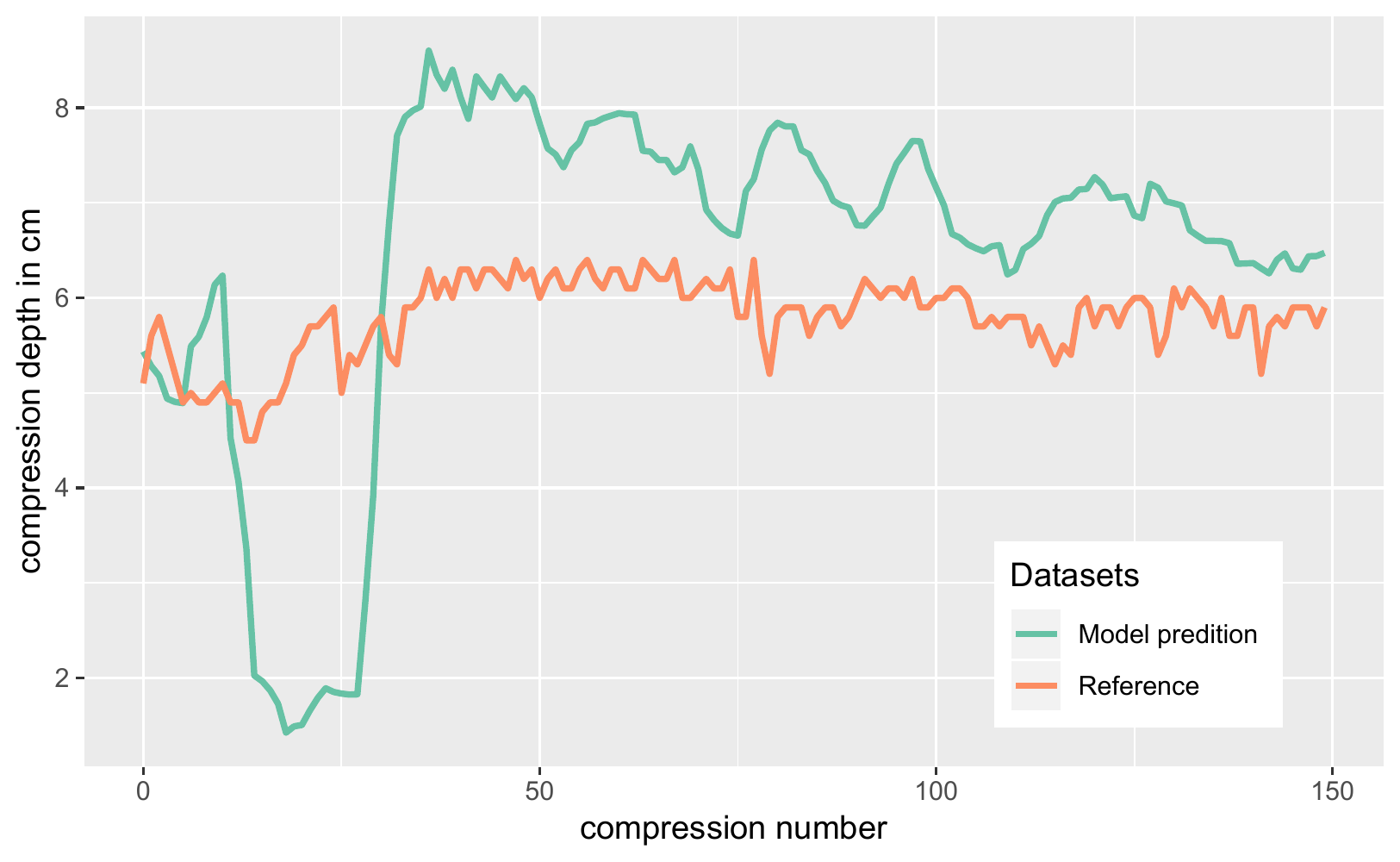}
    \includegraphics[width=7.75cm]{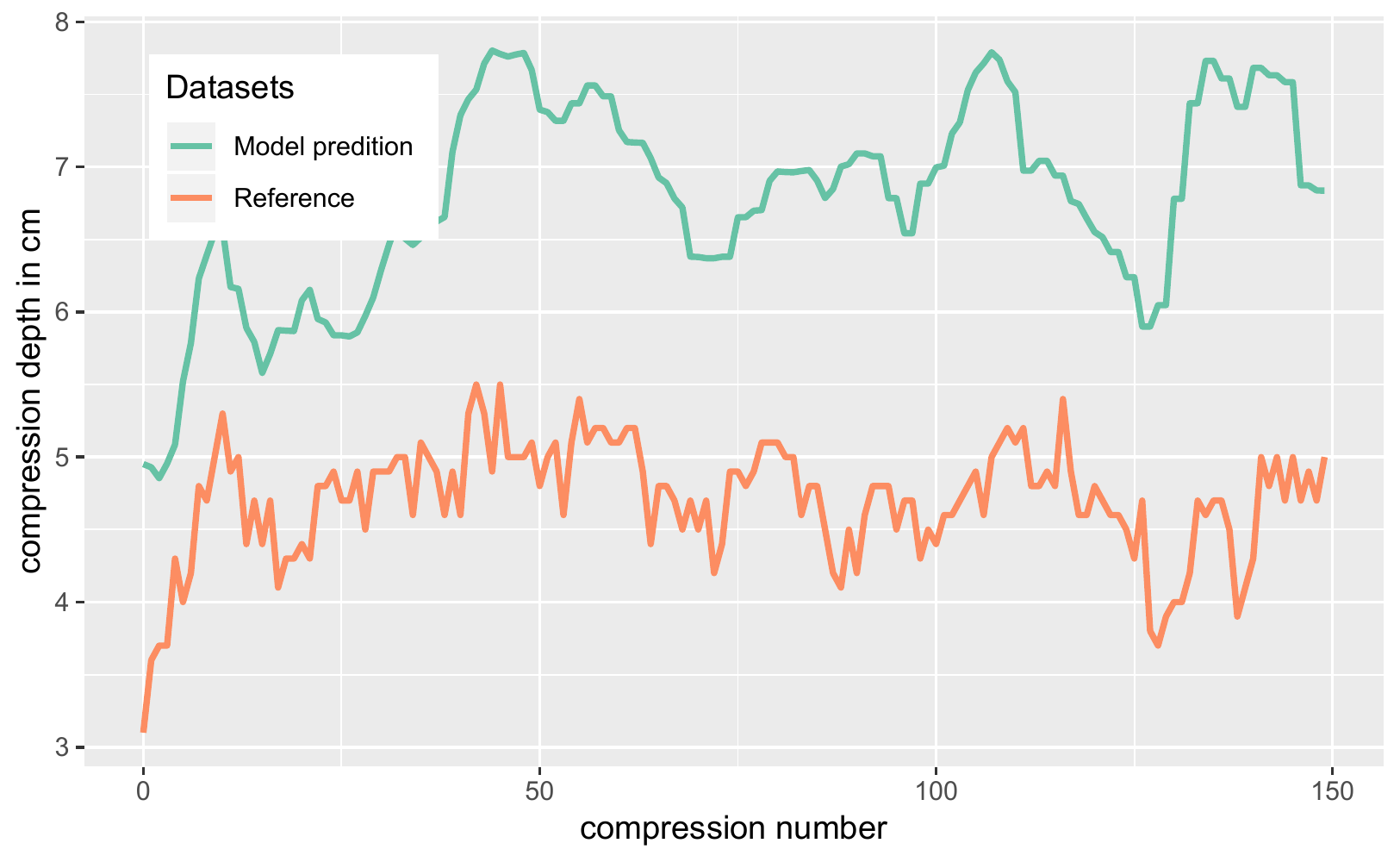}
    \caption{Wrist-worn sensor best-case (left, participant 1) and worst-case (right, participant 12) compression depth prediction.}
    \label{fig:results}
\end{figure}

\FloatBarrier

\subsection{Discussion}
\noindent
In this study, the following two research questions have been investigated:
\begin{enumerate}
\item	How suitable are smartwatch-like inertial sensors for the online detection of the chest compression during CPR regarding the accuracy of resulting CCF and CCD parameters in comparison to the Resusci Anne and the typical hand-holding as reference systems?
\item	How applicable is the DE algorithm for curve-fitting of the chest compression during CPR regarding the accuracy of resulting CCF and CCD parameters for the considered two sensor placements in comparison to the Resusci Anne as reference system?
\end{enumerate}

Regarding the suitability of a smartwatch-like wrist-worn IMU sensor to derive CPR parameters (the first research question), the following promising findings have been made.
The wrist sensor was with an error of 2.0 cpm lightly more accurate regarding CCF compared to an error of 2.1 cpm for the common grasp-in-hand use. Both errors remain in the same order of magnitude – confirming the general robustness of the DE algorithm. Consequently, we could confirm the high suitability of the smartwatch-related wrist positioning for CCF calculation.
The error for the CCD prediction for the wrist sensor is with 1.0 cm  considerably higher than the one of the common grasp-in-hand use (with 0.5 cm). This increased error might be a consequence of the wrist-rotations and the non-orthogonal pressure-distribution during the cardiac massage. Nevertheless, even with the relatively high error of the wrist sensor, a fundamental qualitative statement can be made about the compression depth, since relevant deviations from the optimal compression depth of 5.5 cm (e.g. too low compression depths of 3 cm) are still detectable.

With regard to the second research question, the results show that the use of the DE algorithm is a suitable alternative.  With an error of 2.0 cpm in the prediction of the CCF, this is approximately in the range reported in the literature. Even though Ruiz de Gauna et al. \cite{DeGauna2016} reported a lower median error of 0.9 cpm, the differences can be explained by the different reference system (photoelectric sensor) and another IMU sensor (ADXL330, Analog Devices, USA). In any case, the error at 2.0 cpm is in such a low range that it does not affect the practical application. The target range of $110 \pm10$ cpm can be easily detected. 

In comparison to the related approaches, the achieved CCF accuracies are in within the magnitude of the other approaches and well within the requirements of $\pm10$ cpm of the ERC guidelines \cite{Perkins2015}. In contrast, the CCD results are inconclusive: while the results of some participants are promising, there is much variance in the results (see Figure 6 for a comparison of best-case and worst-case). However, it might be possible to reduce the error further by taking the tilt of the accelerometer into account and apply noise and outlier filters.
Consequently, we could confirm that with wrist-worn devices sufficiently accurate predicting the CCF and CCD. Smartwatches are a well suited unobtrusive and high available alternative platform for giving CPR feedback for bystanders in emergency situations. 

\FloatBarrier

\section{Conclusion}
\noindent
We presented an approach to use sensor data from a wrist-worn IMU and the Differential Evolution (DE) optimization algorithm to dynamically fit a sinusoidal model that can predict frequency and depth parameters for cardiopulmonary resuscitation training or in realistic environments. 
We evaluated the approach with 15 different participants and tested the IMU placement at the wrist and hand for its suitability to derive the parameters. 
The feasibility of the sinusoidal model created from accelerometer data using DE algorithm was confirmed. The chest compression frequency (CCF) could be predicted with a median error of 2.0 cpm and the compression depth (CCD) with a median error of 1.0 cm. Although the CCD leaves room for further improvement, both CCF and CCD can be predicted with sufficient sensitivity for online feedback applications.
Thus, our work represents an initial step towards complete and precise modeling of the CPR using mobile sensors. While focusing on the algorithmic aspects of the detection of the CPR parameters, the general feasibility of smartwatches for CPR feedback (e.g. via a full-featured smartwatch application) has further to be investigated. The development of a corresponding app and its usability studies must be investigated in a subsequent study.

\section*{Acknowledgements}
\noindent
The authors like to thank Max Pfingsthorn for his helpful comments.
This work was supported by the funding initiative Niedersächsisches Vorab of the Volkswagen Foundation and the Ministry of Science and Culture of Lower Saxony as a part of the Interdisciplinary Research Centre on Critical Systems Engineering for Socio-Technical Systems II. 

%The authors would like to thank the anonymous reviewers for their helpful comments.

%\balance
%\vfill

\section*{References}
\bibliographystyle{elsarticle-num}
\bibliography{cpr_imu}

%%%%%%%%%%%%%%%%%%%%%%%%%%%%%%%%%%%%%%%%%%
%% optional
%\sampleavailability{The authors provide the recorded raw sensor data as well as the further processed data, and evaluation scripts together with detailed results for the individual participants.}

%% for journal Sci
%\reviewreports{\\
%Reviewer 1 comments and authors’ response\\
%Reviewer 2 comments and authors’ response\\
%Reviewer 3 comments and authors’ response
%}

%%%%%%%%%%%%%%%%%%%%%%%%%%%%%%%%%%%%%%%%%%
\end{document}